\documentclass[aps,prd,twocolumn,superscriptaddress,nofootinbib,floatfix,10pt]{revtex4-2}

\usepackage{amsmath,amsfonts,amssymb}
\usepackage{bm}
\usepackage{booktabs,etoolbox}
\usepackage{graphicx}
\usepackage{xcolor}
\usepackage[utf8]{inputenc}
\usepackage{orcidlink}
\usepackage{hyperref}
\hypersetup{colorlinks=true,linkcolor=blue,citecolor=blue,urlcolor=blue}
\newcommand{\dd}{\mathrm{d}}

\newcommand{\Mext}{M_{\rm ext}}
\newcommand{\rext}{r_{\rm ext}}
\newcommand{\Ms}{M_{\rm s}}
\newcommand{\Mu}{M_{\rm u}}
\newcommand{\Mq}{M_{\rm q}}
\newcommand{\MPl}{M_{\rm Pl}}
\newcommand{\lPl}{\ell_{\rm Pl}}
\newcommand{\Th}{T_{\rm H}}
\newcommand{\kB}{k_{\rm B}}
\newcommand{\sabs}{\sigma_{\rm abs}}
\newcommand{\lbar}{\bar\lambda}

\begin{document}

\title{Planck-scale black hole remnants inside neutron stars}

\author{Kimet Jusufi\,\orcidlink{0000-0003-0527-4177}}
\email{kimet.jusufi@unite.edu.mk}
\affiliation{Physics Department, State University of Tetovo, Ilinden Street nn, 1200 Tetovo, North Macedonia}

\author{Francisco S. N. Lobo\,\orcidlink{0000-0002-9388-8373}}
\email{fslobo@ciencias.ulisboa.pt}
\affiliation{Instituto de Astrof\'isica e Ci\^encias do Espa\c{c}o, Faculdade de Ci\^encias da Universidade de Lisboa, Edif\'icio C8, Campo Grande, P-1749-016 Lisbon, Portugal}
\affiliation{Departamento de F\'isica, Faculdade de Ci\^encias da Universidade de Lisboa, Edif\'icio C8, Campo Grande, P-1749-016 Lisbon, Portugal}

\begin{abstract}
We examine a Planck-scale regular black-hole remnant in neutron-star matter using a T-duality-inspired effective geometry and explicitly separate geometrical results from assumptions about absorption and emission. The neutral black-hole branch ends at $r_{\rm ext}=\sqrt2\,\ell_0$ and $M_{\rm ext}=(3\sqrt3/4)c^2\ell_0/G$. Within the static family, the surface-gravity temperature grows as $(M-M_{\rm ext})^{1/2}$ after a small mass increase. A constant-greybody Stefan--Boltzmann approximation gives a Hawking power proportional to $(M-M_{\rm ext})^2$ and can produce a stable lower balance point and an unstable growth threshold under stated hypotheses. For $\ell_0=\ell_{\rm Pl}$, however, the formal lower equilibrium has an excess mass of only $10^{-45}$--$10^{-46}\,\mathrm{kg}$ for order-unity absorption efficiency, so the canonical description is inconsistent with individual neutron captures. For a fiducial area-scale absorption cross section with unit coefficient, the mean capture interval is $5.4\times10^8\,\mathrm{yr}$ and the initial nonradiating growth timescale is $9.1\times10^{27}\,\mathrm{yr}$. Under this assumption, a Planck-mass remnant inside a neutron star is therefore effectively inert and, without additional feeding, cannot destroy the star within the age of the Universe. A stationary capture--relaxation state additionally requires an effective microscopic emission law; it is not established by zero temperature alone. Hydrodynamic accretion requires control of wave effects and a mean-free-path criterion; its onset corresponds to masses of order $10^{11}$--$10^{12}\,\mathrm{kg}$ for representative matter parameters. Direct zero-point-length corrections there are negligible, of order $10^{-40}$ at a femtometre. We discuss conditional growth thresholds, critical slowing near an unstable balance, dark-matter feeding, angular-momentum transfer, and the requirements for neutron-star survival constraints. Non-evaporation does not by itself imply stellar destruction, but quantitative predictions require the remnant absorption and relaxation physics.
\end{abstract}
\maketitle

% =================================================
\section{Introduction}
\label{sec:intro}

The possibility that a neutron star hosts a much smaller black hole connects compact-star physics, dark matter, and black-hole thermodynamics. Primordial black-hole capture has been studied as a possible source of central seeds~\cite{Capela2013}, while captured asymmetric dark matter may thermalize, self-gravitate, and collapse in the stellar core~\cite{Goldman1989,Kouvaris2011}. Constraints from stellar survival require both a credible seed-production or capture probability and sufficiently rapid subsequent growth. They are therefore sensitive to the dark-matter environment, microscopic interactions, and the stellar evolution model.

General-relativistic calculations support Bondi-like growth of sufficiently macroscopic central black holes~\cite{Richards2021,Baumgarte2021}. For stiff nuclear equations of state, the relativistic accretion coefficient must be determined consistently~\cite{RichardsBondi2021,RichardsErratum2021}; piecewise-polytropic treatments provide a more realistic description of cold nuclear matter~\cite{Schnauck2021}. Smaller black holes instead require quantum absorption and a check of the neutron mean free path~\cite{Giffin2022}. A recent treatment combines baryonic accretion, semiclassical Hawking evaporation and sustained dark-matter feeding~\cite{Adarsha2026}. Such evolution models employ a Schwarzschild evaporation law without the finite-mass endpoint considered here. Complete disappearance at the quantum-gravity endpoint is an additional assumption, not a consequence established by the semiclassical Hawking calculation alone~\cite{Hawking1975}.

A fundamental zero-point length can motivate a regular effective source through path-integral duality and string T-duality~\cite{Padmanabhan1997,Fontanini2006}. The neutral effective metric of Nicolini, Spallucci and Wondrak has a cold extremal endpoint and a positive-heat-capacity branch~\cite{Nicolini2019}; its geodesic completeness has been examined in Ref.~\cite{Jusufi:2024dtr}, and related quantum-gravity-inspired regularisations have been applied to other spacetimes, such as warp-drive geometries~\cite{Jusufi:2026jan}. These are properties of a specified stationary family; they do not determine its microscopic state spectrum or prove dynamical stability. Broader remnant proposals and their conceptual difficulties are reviewed in Ref.~\cite{Chen2015}. 

Regular remnants with de Sitter interiors have also been considered as dark matter, including possible neutron-star effects through different microscopic mechanisms~\cite{Dymnikova2015}; regular black holes with gravitational self-energy have likewise been proposed as dark matter~\cite{Jusufi:2025qgd}. The general issue of accretion by hypothetical stable microscopic black holes has important precedents~\cite{Giddings2008}; the present contribution focuses on the finite-mass extremal endpoint and the distinction between formal thermal balance and individual capture events.
Related applications include quantum-inspired wormhole constructions~\cite{Lobo:2025nng} and the linearized radial stability analysis of T-duality-inspired thin-shell wormholes~\cite{Lobo:2026dft}.

In a charged extension of the T-duality-inspired geometry, a sufficiently near-endpoint part of the extremal branch avoids a local charged-scalar Schwinger instability for fixed massive probe species~\cite{Chen2026}. This result concerns a specific charged probe and does not establish neutral relaxation, stability under matter accretion, or persistence of a regular interior. The latter questions require a dynamical completion~\cite{Carballo2018}.

We ask whether non-evaporation alone makes a Planck-scale remnant an efficient seed of stellar destruction. We show that this inference does not follow from the cold endpoint. Within an explicitly stated area-scale absorption model, initial baryonic growth is extremely slow even if the absorbed energy is retained. A lower attracting balance is possible in a continuous emission model, but its canonical estimate lies outside that model's domain at the Planck scale. The physical alternatives are stochastic relaxation, slow accumulation, or modified capture, depending on microscopic transition rates. The large separation between the extremal scale and nuclear accretion scales survives this distinction; an actual two-root phase diagram does not follow without further assumptions.

The paper is organised in the following manner: 
Section~\ref{sec:geometry} establishes the effective geometry, its near-extremal thermodynamics and the limits of a canonical description. Section~\ref{sec:accretion} discusses discrete capture, the kinetic-to-fluid crossover and relativistic accretion. Section~\ref{sec:master} develops conditional balance points, feeding and threshold-crossing times. Section~\ref{sec:discussion} explains the physical limitations and observational requirements. Section~\ref{sec:conclusions} gives the conclusions.

Throughout, $G$ denotes Newton's gravitational constant, $M$ is a physical mass, and $\ell_0$ is a length. We retain the required factors of $G$, $c$, $\hbar$, and $k_B$ explicitly.

% =================================================
\section{Effective geometry and near-extremal thermodynamics}
\label{sec:geometry}

\subsection{Regular geometry and cold endpoint}

We consider the static, spherically symmetric line element
\begin{equation}
ds^2=-f(r)\,c^2dt^2+\frac{dr^2}{f(r)}+r^2d\Omega^2,
\end{equation}
with \cite{Nicolini2019}
\begin{equation}
f(r)
=
1-
\frac{2GM}{c^2}
\frac{r^2}{(r^2+\ell_0^2)^{3/2}}.
\label{metricfunction}
\end{equation}
Interpreted through the Einstein equations, this geometry is sourced by the effective mass density \cite{Nicolini2019}
\begin{equation}
\rho_{\rm eff}(r)
=
\frac{3M\ell_0^2}{4\pi(r^2+\ell_0^2)^{5/2}},
\label{density}
\end{equation}
whose enclosed mass is $m(r)=Mr^3/(r^2+\ell_0^2)^{3/2}$, so that $f=1-2Gm(r)/(c^2r)$. The spacetime is regular at $r=0$, where it is locally de Sitter, and approaches Schwarzschild for $r\gg\ell_0$ \cite{Nicolini2019}.

The reciprocal lapse and radial metric coefficient impose an anisotropic effective stress, rather than an ordinary neutron fluid:
\begin{equation}
 p_{r,\rm eff}=-\rho_{\rm eff}c^2,\qquad
 p_{t,\rm eff}=\rho_{\rm eff}c^2
 \frac{3r^2-2\ell_0^2}{2(r^2+\ell_0^2)}.
 \label{effectivepressures}
\end{equation}
Here $\rho_b$ will denote the separate stellar baryon rest-mass density. Changing $M$ while keeping $\ell_0$ fixed is a quasistatic prescription for a family of effective sources, not a derived time-dependent solution of a covariant microscopic theory.

The horizon condition $f(r_+)=0$ gives the mass--radius relation
\begin{equation}
M(r_+)
=
\frac{c^2}{2G}
\frac{(r_+^2+\ell_0^2)^{3/2}}{r_+^2}.
\label{MassRadius}
\end{equation}
Since $d\ln M/dr_+=(r_+^2-2\ell_0^2)/[r_+(r_+^2+\ell_0^2)]$, the minimum occurs at $\rext=\sqrt2\,\ell_0$, where \cite{Nicolini2019}
\begin{equation}
\Mext
=
\frac{3\sqrt3}{4}\frac{c^2\ell_0}{G}.
\label{Mext}
\end{equation}
For $M<\Mext$ there is no horizon, while for $M>\Mext$ there is an inner and an outer horizon; below we always denote by $r_+$ the outer one. For $\ell_0=\lPl=\sqrt{\hbar G/c^3}$, with $\MPl=\sqrt{\hbar c/G}$, one has $c^2\ell_0/G=\MPl$, so that $\Mext\simeq1.30\,\MPl\simeq2.8\times10^{-8}\,\mathrm{kg}$ and $\rext\simeq2.3\times10^{-35}\,\mathrm{m}$.

The Hawking temperature, $\kB\Th=\hbar c\,f'(r_+)/(4\pi)$, is
\begin{align}
\kB\Th
&=
\frac{\hbar c}{4\pi r_+}
\left(
1-\frac{3\ell_0^2}{r_+^2+\ell_0^2}
\right)
=
\frac{\hbar c}{4\pi}
\frac{r_+^2-2\ell_0^2}
{r_+(r_+^2+\ell_0^2)}.
\label{TH}
\end{align}
The horizon condition in Eq.~\eqref{MassRadius} eliminates the explicit factor $GM/c^2$ from Eq.~\eqref{TH}; for $r_+\gg\ell_0$, the result reduces to $k_BT_H\simeq\hbar c^3/(8\pi GM)$.
Consequently, $\Th(\rext)=0$: the stationary black-hole branch has a cold endpoint. Whether evaporation approaches or reaches this endpoint is a question about the emission law. The temperature is maximal where $r_+^4-7\ell_0^2r_+^2-2\ell_0^4=0$, i.e.\ at $r_+=\big[(7+\sqrt{57})/2\big]^{1/2}\ell_0\simeq2.70\,\ell_0$, corresponding to $M\simeq1.64\,c^2\ell_0/G$ and
\begin{equation}
\kB T_{\rm max}\simeq0.0188\,\frac{\hbar c}{\ell_0}.
\label{Tmax}
\end{equation}
For $\ell_0=\lPl$ this is $\kB T_{\rm max}\simeq2.3\times10^{17}\,\mathrm{GeV}$. The heat capacity $C_H=d(Mc^2)/d\Th$ is positive on the small branch $\Mext<M<M(T_{\rm max})$ and negative on the Schwarzschild-like branch. The positive heat capacity characterizes the response within this static family; it is not a proof of dynamical or quantum stability. At the endpoint, the central Ricci scalar is $R(0)=24G\Mext/(c^2\ell_0^3)=18\sqrt3/\ell_0^2$. For $\ell_0=\lPl$, the curvature is Planckian, so the classical metric and surface-gravity temperature are effective extrapolations.

\subsection{Near-extremal expansion and an illustrative emission law}
\label{sec:nearext}

Within the neutral, nonrotating stationary family, the surface-gravity temperature vanishes only at precisely $M=\Mext$. If it absorbs an energy $\delta M\,c^2$, then $M=\Mext+\delta M$ and the object is no longer extremal. To quantify this, define the dimensionless variables
\begin{equation}
x=\frac{r_+}{\ell_0},
\qquad
\mu=\frac{GM}{c^2\ell_0},
\end{equation}
in terms of which Eq.~\eqref{MassRadius} reads
\begin{equation}
\mu(x)
=
\frac{(x^2+1)^{3/2}}{2x^2}.
\end{equation}
Expanding around the extremal point $x_0=\sqrt2$, $\mu_0=3\sqrt3/4$, and using $\mu^{\prime\prime}(x_0)=2\mu_0/3=\sqrt3/2$, one obtains
\begin{equation}
\mu-\mu_0
=
\frac{\sqrt3}{4}(x-\sqrt2)^2+\mathcal{O}\!\left[(x-\sqrt2)^3\right].
\label{nearExtMass}
\end{equation}
Near extremality Eq.~\eqref{TH} gives $\kB\Th\simeq\hbar c\,(x-\sqrt2)/(6\pi\ell_0)$, and therefore
\begin{equation}
\kB\Th
\simeq
\frac{\hbar c}{3^{5/4}\pi\,\ell_0}
\sqrt{
\frac{G\,\delta M}{c^2\ell_0}
},
\qquad
\Th\propto\sqrt{M-\Mext}.
\label{nearExtTemperature}
\end{equation}
The geometrical temperature therefore rises after a small mass increase along this family. A physical excitation need not have a canonical thermal spectrum, and capture can also change angular momentum or charge.

For illustration, adopt a semiclassical Stefan--Boltzmann approximation with a constant effective factor $g_\ast$ representing the assumed emissivity and particle degrees of freedom:
\begin{equation}
P_{\rm H}\simeq g_\ast\frac{\pi^2}{120}\frac{A_+(\kB\Th)^4}{\hbar^3c^2},
\label{PH}
\end{equation}
where $A_+=4\pi r_+^2\simeq8\pi\ell_0^2$ near extremality. Hence
\begin{equation}
P_{\rm H}\propto(M-\Mext)^2
\label{PHNearExt}
\end{equation}
in this approximation close to $\Mext$. Actual spin-dependent greybody factors and available emission channels may change the power law~\cite{Page1976,Bai2023}.

More explicitly, the corresponding mass-loss coefficient is
\begin{equation}
 \dot M_H\simeq B_H(\delta M)^2,\qquad
 B_H=\frac{g_\ast\hbar G^2}
 {15\,3^5\pi c^4\ell_0^4}.
 \label{BHnear}
\end{equation}
For constant $g_\ast$, the full approximation has
\begin{equation}
 P_H(x)\propto\frac{(x^2-2)^4}{x^2(x^2+1)^4},
 \qquad
 x_{P}=\sqrt{\frac{13+\sqrt{177}}{2}}\simeq3.63.
 \label{powermaximum}
\end{equation}
Thus power peaks at $M\simeq2.024\,c^2\ell_0/G$, later than the temperature maximum: area growth must be included when discussing the turnover of Hawking loss. These maxima describe the approximation, not a controlled Planck-scale emission prediction. Recent mode-resolved greybody calculations in a different quantum-corrected Schwarzschild geometry likewise illustrate that late-time power depends on the transmission spectrum~\cite{Shafiee2026}; their quantitative rates cannot be transferred to the present effective metric.

\subsection{Limits of the canonical description}
\label{sec:canonical}

Equation~\eqref{nearExtTemperature} must be applied with care. Energy conservation forbids the emission of quanta with energy larger than the excess energy $\delta M\,c^2$ above the ground state, whereas a thermal spectrum is dominated by quanta of energy $\sim\kB\Th$. An order-of-magnitude energy-consistency diagnostic for a canonical approximation is $\kB\Th\lesssim\delta M\,c^2$, which, using Eq.~\eqref{nearExtTemperature}, requires
\begin{equation}
\delta M\gtrsim\delta M_\ast
\equiv
\frac{\hbar^2G}{3^{5/2}\pi^2c^4\ell_0^3}
\simeq
6.5\times10^{-3}\left(\frac{\lPl}{\ell_0}\right)^{3}\MPl.
\label{dMstar}
\end{equation}
For $\ell_0=\lPl$ one finds $\delta M_\ast\simeq1.4\times10^{-10}\,\mathrm{kg}$, i.e.\ about $10^{17}$ neutron masses. As an illustration, a single absorbed neutron ($\delta M=m_n$) would formally yield $\kB\Th\simeq3\times10^{8}\,\mathrm{GeV}$, while only $\simeq1\,\mathrm{GeV}$ is available for emission. The scale $\delta M_\ast$ is a diagnostic rather than a sharp boundary of validity: a controlled thermal spectrum requires $\kB\Th\ll\delta M c^2$, as well as negligible backreaction and sufficiently small statistical fluctuations~\cite{Preskill1991}. At a single-neutron excitation, the canonical spectrum is energetically inconsistent. Its lower balance point below is consequently a diagnostic of model failure, rather than a quantitative equilibrium estimate.

An energy-conserving description must specify transition rates between remnant states. If $\gamma(E;M)\,\dd E$ is the emission probability per unit time into quanta with energies in $[E,E+\dd E]$, a schematic microscopic power is
\begin{equation}
 P_{\rm mic}(M)=\int_0^{(M-\Mext)c^2}
 E\,\gamma(E;M)\,\dd E,
 \label{microPower}
\end{equation}
where the upper limit is an energy-conservation bound for a neutral ground state at $\Mext$; recoil and selection rules can lower the actual kinematic endpoint, and spin- or charge-dependent final states require their own energy bounds. Energy conservation alone does not supply $\gamma$. Density-of-states factors, matrix elements and transmission probabilities are additional inputs. Microcanonical and energy-conserving approaches provide precedents~\cite{Casadio1998,Parikh2000}, but no such transition law has been derived here.

Recent quantum calculations of charged near-extremal black holes show that backreaction, fluctuations of the near-horizon geometry and angular-momentum constraints can strongly modify emission~\cite{Brown2026}. The T-duality-inspired neutral endpoint has a classical ${\rm AdS}_2\times S^2$ throat with radii $\sqrt3\,\ell_0$ and $\sqrt2\,\ell_0$, but this resemblance does not justify importing a charged black hole's state spectrum or emission law. Establishing an analogous quantum description requires the underlying action and state space.

% =================================================
\section{Microscopic capture and the transition to fluid accretion}
\label{sec:accretion}

\subsection{Absorption ansatz and initial growth}
\label{sec:quantum}

When wave effects or the neutron mean free path invalidate hydrodynamics, the Bondi formula cannot be extrapolated. A useful phenomenological description is
\begin{equation}
\dot M_q
\simeq
\rho_b\,v_F\,
\sabs(M,\ell_0),
\label{QuantumAccretion}
\end{equation}
where this rest-mass estimate uses a representative speed, rather than the full incident distribution. The cross section must ultimately follow from the absorption probability of neutron or quark modes. As an explicitly phenomenological fiducial estimate we write
\begin{equation}
\sabs
=
\mathcal{C}(\omega,M,\ell_0)\,\pi r_+^2,
\label{SigmaAbs}
\end{equation}
where $\omega$ denotes the incident mode frequency and $\mathcal C$ encodes spin, velocity and geometry dependence. A scalar horizon-area absorption result is not a massive-neutron result~\cite{Das1997}. Schwarzschild calculations provide a useful Dirac-field methodology~\cite{Unruh1976,Doran2005}, but do not establish the cross section at an extremal regular horizon. Quantum corrections can also alter near-extremal scalar absorption~\cite{Emparan2025}; those results cannot be quantitatively transferred to neutrons in this effective geometry.

More generally, with $\langle\cdot\rangle_F$ an average over the occupied stellar neutron momentum distribution,
\begin{equation}
 \Gamma_{\rm cap}=n_b\langle\sabs v\rangle_F,\qquad
 \dot M_b=n_b\left\langle\frac{\mathcal E}{c^2}\sabs v\right\rangle_F.
 \label{captureaverage}
\end{equation}
Here $\mathcal E$ is the energy per captured quasiparticle referred to the matching clock used for the black-hole mass. Rest mass and kinetic or interaction contributions must be distinguished in a quantitative nuclear-matter model. In the estimates below we use $\mathcal E\simeq m_nc^2$, $v_F=0.4c$, and an effective distribution-averaged $\mathcal C$; these are illustrative assumptions. All numerical rates and times use the locally normalized clock at a matching radius outside the small hole and well inside the stellar pressure scale. If $\dd\tau_{\rm match}=N_\star\dd t_\infty$, where $t_\infty$ is the exterior stellar Killing time, a quoted local interval corresponds to $t_\infty=\tau_{\rm match}/N_\star$ for a stationary background. The stellar lapse $N_\star$ is not calculated here. Comparisons with observed stellar ages must include this conversion; the stated local estimates do not specify a global remnant-plus-star solution.

At extremality $r_+=\sqrt2\,\ell_0$, so for $\ell_0=\lPl$, we get
\begin{equation}
\sabs
\simeq
2\pi\mathcal{C}\,\lPl^2
\simeq
1.6\times10^{-69}\,\mathcal{C}\ {\rm m^2}.
\end{equation}
With $\rho_b\sim5\times10^{17}\,{\rm kg\,m^{-3}}$ and $v_F\simeq0.4\,c$ one obtains
\begin{equation}
\dot M_q
\sim
9.84\times10^{-44}\,\mathcal{C}\ {\rm kg\,s^{-1}}.
\label{MdotPlanck}
\end{equation}

Ignoring Hawking loss altogether, an order-unity increase of the remnant mass would require
\begin{equation}
t_{\rm grow}
\sim
\frac{\Mext}{\dot M_q}
\sim
9.10\times10^{27}\,\mathcal{C}^{-1}\ {\rm yr},
\label{GrowthTime}
\end{equation}
i.e.\ $10^{26}$--$10^{28}$ years for $\mathcal{C}\sim1$--$100$.

For a constant $\mathcal C$ area-scale ansatz, the characteristic initial growth time is $t_{\rm grow}=\Mext/\dot M_q(\Mext)$. It is not an exact doubling time because the cross section changes with mass. Direct integration along the outer-horizon branch gives
\begin{align}
 t_2&=\frac{c^2}{\pi\mathcal C\rho_bv_FG\ell_0}
 \int_{\mu_0}^{2\mu_0}\frac{\dd\mu}{x_+(\mu)^2},\nonumber\\
 &\simeq0.226\,t_{\rm grow}
 \simeq2.06\times10^{27}\mathcal C^{-1}\,\mathrm{yr},
 \label{doublingtime}
\end{align}
where $\mu_0=3\sqrt3/4$ and $x_+(\mu)$ is the outer solution of the mass--radius relation. For this ansatz, the time to much larger masses is dominated by the initial stage and remains of order $t_{\rm grow}$. In the actual quantum regime $\mathcal C$ need not be constant: the Unruh-based rate can cross from $M^2$ to an approximate $M^3$ law~\cite{Giffin2022}. The slow-growth conclusion is therefore conditional on the initial area-scale absorption assumption and the absence of an additional mass supply or a new absorption channel.

\subsection{Discrete capture and possible relaxation}

\label{sec:discrete}
The assumed extremal capture rate corresponds to
\begin{equation}
 t_{\rm cap}=\Gamma_{\rm cap}^{-1}
 \simeq\frac{m_n}{\dot M_q}
 \simeq5.39\times10^8\mathcal C^{-1}\,\mathrm{yr}.
 \label{tcap}
\end{equation}
It implies approximately $19\mathcal C$ captures over $10^{10}$ years of matching-frame time. The interval is not a prediction of the effective metric alone: it inherits the cross-section assumption and the rest-mass approximation of Eq.~\eqref{captureaverage}.

A rare capture raises the mass by $\Delta M=\mathcal E/c^2$, of order one neutron mass for the chosen incident states. It may then relax, retain the excitation, or access other internal channels. A return to a remnant-like state requires an allowed microscopic emission process, and spending most of the time close to the ground state requires sufficiently rapid effective relaxation relative to Eq.~\eqref{tcap}. Neither condition follows from $\Th(\Mext)=0$.

For well-separated excitations with an integrable excess-mass profile $\delta M_{\rm exc}(t)$, the leading rare-event estimate is
\begin{align}
 \langle M-\Mext\rangle
 &\simeq\Gamma_{\rm cap}\int_0^\infty
 \delta M_{\rm exc}(t)\,\dd t,\nonumber\\
 &=\Delta M\,\frac{\tau_{\rm eff}}{t_{\rm cap}},\nonumber\\
 \tau_{\rm eff}&\equiv\frac{1}{\Delta M}
 \int_0^\infty\delta M_{\rm exc}(t)\,\dd t.
 \label{averageMs}
\end{align}
For exponential relaxation, $\tau_{\rm eff}$ is the decay time. For an algebraic tail the integral may not exist; for example, extrapolating $\dot{\delta M}=-B_H\delta M^2$ gives a logarithmically divergent isolated-excitation integral. That extrapolation is not physically justified in this regime, but shows why a generic ``relaxation time'' cannot replace an emission law. Overlapping events or long tails require a stochastic master equation; its mean drift involves the average of the loss function, not the loss evaluated at the mean mass.

If the absorbed energy is retained, Eqs.~\eqref{GrowthTime} and \eqref{doublingtime} still imply very slow initial growth under the same capture model. Non-evaporation therefore does not logically imply rapid stellar consumption, even without invoking an attracting microscopic state. This conditional inference is distinct from a proof that all possible remnant absorption laws are harmless.

\subsection{Wavelength and transport criteria}
\label{sec:fluid}

Hydrodynamics requires a scale hierarchy: wave effects must be controlled, and the collision or relaxation length must be sufficiently shorter than the scale of inflow. The characteristic gravitational accretion scale is the Bondi radius
\begin{equation}
r_B
\sim
\frac{GM}{c_s^2},
\label{BondiRadius}
\end{equation}
where $c_s$ is the sound speed of nuclear matter, whereas the characteristic microscopic scale of the medium is the reduced neutron Fermi wavelength
\begin{equation}
\bar\lambda_F=\frac{1}{k_F}
=
(3\pi^2 n_b)^{-1/3}.
\label{FermiWavelength}
\end{equation}
For a central density $\rho_b\sim5\times10^{17}\,{\rm kg\,m^{-3}}$ one has $n_b\simeq\rho_b/m_n\simeq3\times10^{44}\,{\rm m^{-3}}$, giving $\lbar_F\simeq5\times10^{-16}\,{\rm m}$, a Fermi momentum $p_Fc\simeq0.4\,\mathrm{GeV}$, and a Fermi velocity $v_F\simeq0.4\,c$.

A wavelength-based indicator, which by itself does not establish hydrodynamics, is $r_B\gtrsim\lbar_F$, i.e.\ for
\begin{equation}
M\gtrsim M_\lambda
\sim
\frac{c_s^2\lbar_F}{G}.
\label{Mq}
\end{equation}
For $c_s\simeq(0.3$--$0.7)\,c$ this gives
\begin{equation}
M_\lambda\sim
6\times10^{10}
\text{--}
3\times10^{11}\ {\rm kg},
\label{MqValue}
\end{equation}

The length-based indicator is already about nineteen orders of magnitude above the Planck-scale extremal mass. This estimate is a comparison of lengths, not a validated fluid-transition mass. In particular, it must be supplemented by the mean-free-path condition discussed in Ref.~\cite{Giffin2022}.

If $\lambda_{\rm mfp}=1/(n_b\sigma_{nn})$ is an effective collision length, a kinetic condition is $r_B\gg\lambda_{\rm mfp}$, or equivalently a small Knudsen number. At the onset $r_B\sim\lambda_{\rm mfp}$, kinetic corrections are generally substantial. For illustration, $n_b\simeq0.3\,\mathrm{fm^{-3}}$ and $\sigma_{nn}\simeq10\,\mathrm{mb}$ give $\lambda_{\rm mfp}\simeq3.3\,\mathrm{fm}$. Using the convention $R_B=2GM/c_s^2$ and $c_s^2=0.3c^2$ adopted in Ref.~\cite{Giffin2022} yields an onset near $6.7\times10^{11}\,\mathrm{kg}$. Factors of order unity change with radius convention; medium effects, Pauli-limited collisions and superfluidity can change the relaxation length more substantially. Representative crossover scales are therefore of order $10^{11}$--$10^{12}\,\mathrm{kg}$, rather than a universal $\Mq$ fixed by the Fermi wavelength. A collisionless classical-capture interval may also intervene. Table~\ref{tab:scales} summarizes these crossover estimates alongside the remnant parameters and the capture, growth, and evaporation scales derived above.

\subsection{Local relativistic fluid equations}
\label{sec:bondi}

When the fluid conditions hold, local test-fluid accretion can be formulated in the effective metric~\cite{Bondi1952,Michel1972}. We retain $c$ explicitly and define $u^\mu=dx^\mu/d\tau$, with $\tau$ the proper time, and the dimensionless radial four-velocity component $w=u^r/c=(dr/d\tau)/c$. For stationary, spherically symmetric flow of a perfect fluid with particle-number density $n$ and four-velocity $u^\mu$, particle-number conservation $\nabla_\mu(nu^\mu)=0$ gives
\begin{equation}
4\pi r^2 n c w
=
{\rm const},
\label{Continuity}
\end{equation}
and for isentropic flow the relativistic Bernoulli relation reads
\begin{equation}
h\sqrt{f+w^2}
=
{\rm const},
\label{Bernoulli}
\end{equation}
where $h=(\epsilon+P)/n$ is the enthalpy per particle. Here $\epsilon$ is the fluid energy density, $P$ is its pressure, and $n$ is its proper particle-number density; $h$ therefore has units of energy. For an isentropic barotropic fluid with $a^2=\dd P/\dd\epsilon$ in a metric of the form $ds^2=-f\,c^2dt^2+f^{-1}dr^2+r^2d\Omega^2$, the candidate regularity conditions at the sonic point $r_c$ are \cite{Michel1972}
\begin{equation}
w_c^2
=
\frac{r_cf'(r_c)}{4},
\qquad
a_c^2
=
\frac{r_cf'(r_c)}
{4f(r_c)+r_cf'(r_c)},
\label{Sonic}
\end{equation}
where $a=c_s/c$ is the dimensionless sound speed, evaluated at the sonic point when subscripted by $c$. For the T-duality geometry
\begin{equation}
f'(r)
=
\frac{2GM}{c^2}
\frac{r(r^2-2\ell_0^2)}
{(r^2+\ell_0^2)^{5/2}},
\end{equation}
so that
\begin{equation}
w_c^2
=
\frac{GM}{2c^2}
\frac{r_c^2(r_c^2-2\ell_0^2)}
{(r_c^2+\ell_0^2)^{5/2}}.
\label{TdualSonic}
\end{equation}
These are local regularity conditions, not a complete accretion solution. The equation of state and the outer density and enthalpy must select a globally admissible transonic flow with a critical point outside the outer horizon; in particular $w_c^2\geq0$ requires $r_c\geq\sqrt2\ell_0$.

For $r_c\gg\ell_0$ the metric function has the expansion
\begin{equation}
f(r)
=
1-\frac{2GM}{c^2r}
+
\frac{3GM\ell_0^2}{c^2r^3}
+\mathcal{O}\!\left(\frac{GM\ell_0^4}{c^2r^5}\right),
\label{LargeRExpansion}
\end{equation}
so the relative correction to the Newtonian potential is $\delta\Phi/\Phi\simeq-3\ell_0^2/(2r_c^2)$. For a nuclear-scale sonic radius $r_c\sim10^{-15}\,\mathrm m$ and $\ell_0=\lPl$, the ratio is $\ell_0^2/r_c^2\simeq2.6\times10^{-40}$; its precise value depends on the selected sonic point. Once ordinary Bondi accretion begins, direct $\ell_0$ corrections are therefore negligible. The role of regularity is to alter the microscopic initial condition and the evaporation endpoint, not the macroscopic Bondi stage.

The neutron-star background is not globally asymptotically flat on the scale of the local inflow. A matched calculation must relate the local accretion clock and enthalpy to the stellar lapse, and cannot impose the isolated reciprocal-metric ansatz on the entire stellar interior. The small-hole approximation is useful because the capture region is tiny relative to the stellar pressure scale, but it is not a self-consistent remnant-plus-star solution.

% =================================================
\section{Conditional balances, feeding and stellar growth}
\label{sec:master}

\subsection{Stationary balance points and their domain}

We model the evolution of the central black-hole mass by
\begin{equation}
\frac{dM}{dt}
=
F(M,t)
\equiv
\dot M_b(M)
+
\dot M_{\rm DM}(M,t)
-
\dot M_{\rm H}(M),
\label{Master}
\end{equation}
where rates are referred to a common matching clock, $\dot M_b$ is the captured baryonic energy flux divided by $c^2$, $\dot M_{\rm DM}$ is an effective dark-matter supply, and $\dot M_H\geq0$ is an emitted-energy loss. A stationary phase portrait requires time-independent feeding and a fixed stellar background; only in that case do we abbreviate $F(M,t)$ by $F(M)$. For a simple root, $F(M_\star)=0$, $F'(M_\star)<0$ implies attraction and $F'(M_\star)>0$ implies repulsion. With time-dependent feeding, moving balance curves are not fixed points; adiabatic tracking requires the forcing time to exceed the local relaxation time.

If the cold ground state does not emit and the absorption rate is nonzero, $F(\Mext)>0$. This positive endpoint drift alone does not imply a nearby stable root. In the illustrative canonical model, assume an approximately constant weak input $A_0=\dot M_b(\Mext)+\dot M_{\rm DM}(\Mext)$ and the coefficient $B_H$ of Eq.~\eqref{BHnear}. Then
\begin{equation}
 F(M)\simeq A_0-B_H(\delta M)^2,
 \qquad \Ms-\Mext\simeq\sqrt{A_0/B_H},
 \label{lowerroot}
\end{equation}
with $F'(\Ms)\simeq-2\sqrt{A_0B_H}<0$ provided the neglected input variation is sufficiently small. For the fiducial baryonic rate and no extra feeding,
\begin{equation}
 \Ms-\Mext\simeq1.15\times10^{-45}
 \sqrt{\mathcal C/g_\ast}\,\mathrm{kg}.
 \label{formalMs}
\end{equation}
For $g_\ast=1$--$100$, this is $10^{-45}$--$10^{-46}\sqrt{\mathcal C}\,\mathrm{kg}$, far below both $m_n$ and $\delta M_\ast$. It is a formal root in an invalid canonical regime and cannot be identified with the mean of a physical stationary state. Such a state instead requires the transition law and rare-event conditions of Sec.~\ref{sec:discrete}.

For a fixed set of emitted particle species, the large-mass Hawking loss scales as $M^{-2}$ while fixed-background Bondi accretion grows as $M^2$. The power turnover occurs after the temperature maximum even in the approximation of Eq.~\eqref{powermaximum}. For a continuous stationary $F$, positive drift near the ground state, negative drift somewhere at intermediate mass, and positive drift at large mass imply at least two crossings. A simple positive-to-negative crossing is attracting; a simple negative-to-positive crossing is repelling. Exactly two roots require an additional restriction on the rate functions. If these hypotheses select one lower root $\Ms$ and one upper root $\Mu$, the illustrative phase portrait is
\begin{equation}
\begin{aligned}
\Mext<M<\Ms &:\quad \dot M>0 \quad (\text{approach to } \Ms),\\
\Ms<M<\Mu &:\quad \dot M<0 \quad (\text{relaxation to } \Ms),\\
M>\Mu &:\quad \dot M>0 \quad (\text{runaway growth}).
\end{aligned}
\label{PhasePortrait}
\end{equation}
The lower canonical root is absent from the conventional Schwarzschild rate model, which has no finite-mass cold endpoint. In a remnant model, subthreshold evolution depends on microscopic losses and absorption: return to a near-ground-state distribution is one possible outcome, not a consequence proved by the geometrical temperature.

\subsection{An illustrative upper threshold}
\label{sec:runaway}

For $M\gg\Mext$ the regular black hole approaches Schwarzschild, and its Hawking mass-loss rate can be written as \cite{Page1976,MacGibbon1991}
\begin{equation}
\dot M_{\rm H}
\simeq
\frac{\alpha_{\rm H}}{M^2},
\qquad
\alpha_{\rm H}\simeq5.3\times10^{16}\,\varphi(M)\ {\rm kg^3\,s^{-1}}.
\label{HawkingLargeM}
\end{equation}
The dimensionful coefficient $\alpha_H$ already includes the dependence on $G$: it can be written as $\alpha_H=\chi(M)\hbar c^4/G^2$, with dimensionless $\chi(M)$ incorporating the greybody factors and emitted species. Here $\varphi(M)$ counts the effectively emitted particle species ($\varphi=1$ for photons, gravitons, and three light neutrino species, and $\varphi\sim10$--$15$ once quarks and gluons are emitted). For an illustrative hydrodynamic estimate, write the Newtonian-form rate
\begin{equation}
\dot M_{\rm B}
=
K_B M^2,
\qquad
K_B
=
4\pi\lambda
\frac{\rho_b G^2}{c_s^3},
\label{BondiMassRate}
\end{equation}
with an effective $\lambda$ that must be calibrated to the relativistic equation of state and outer boundary conditions~\cite{RichardsBondi2021,RichardsErratum2021,Schnauck2021}. It is not a universal constant for stiff nuclear matter. Equating the two rates yields the unstable threshold
\begin{equation}
\Mu
\simeq
\left(
\frac{\alpha_{\rm H}}{K_B}
\right)^{1/4}.
\label{Mu}
\end{equation}
For $\rho_b\sim5\times10^{17}\,{\rm kg\,m^{-3}}$, $c_s\simeq(0.3$--$0.7)\,c$, $\lambda\sim1$, and $\varphi\sim1$--$15$,
\begin{equation}
\Mu\sim3.4\times10^{10}\text{--}1.3\times10^{11}\ {\rm kg}.
\label{MuValue}
\end{equation}

This range varies $\varphi$ independently for illustration; a consistent solution uses $\alpha_H(\Mu)$ and the appropriate kinetic accretion law. The threshold generally lies below a well-established hydrodynamic regime. In particular, $27\pi G^2M^2/c^4$ is a null geometric-optics capture cross section, not the precise massive-neutron cross section at $v_F=0.4c$. Full quantum absorption can yield a different critical mass; Ref.~\cite{Giffin2022} finds $\sim1.6\times10^{11}\,\mathrm{kg}$ for its adopted matter model. Sustained dark-matter feeding can alter the threshold far more strongly~\cite{Adarsha2026}. The quoted interval is consequently an approximation estimate rather than a universal destruction threshold.

The scale of a baryon-driven upper threshold is vastly larger than $\Mext$, so direct $\ell_0$ corrections are negligible there for a Planck-scale regulator. The cold endpoint changes the possible fate below threshold; it does not, by itself, fix either the threshold or that fate.

The conclusion can change if additional mass is supplied by dark matter, through the term $\dot M_{\rm DM}$ in Eq.~\eqref{Master}. Sustained dark-matter feeding is known to alter the critical seed mass in endoparasitic-black-hole scenarios \cite{Adarsha2026}. For a regular remnant there are two qualitatively different possibilities.

\subsection{Continuous feeding and collective deposition}

Within a stationary deterministic rate model, a sufficiently weak additional input can shift an existing lower attracting root. If an additive constant input $D$ is used, a sufficient condition to remove an entire negative-drift interval $I$ is
\begin{equation}
 D>\sup_{M\in I}\left[\dot M_H(M)-\dot M_b(M)\right].
 \label{feedingbarrier}
\end{equation}
At a smooth saddle-node boundary the balance satisfies $F=0$ and $F'=0$. For mass-dependent or time-dependent feeding the criterion must be applied to the actual input function and evolution history. The height of the loss barrier cannot be inferred from $T_{\max}$ alone, especially when Planck-scale emission is uncontrolled. Stochastic accumulation or rare bursts can also differ from continuous adiabatic tracking; no universal feeding rate for escape is established here.

Collective deposition is another possibility if dark matter first accumulates into a self-gravitating core and subsequently collapses onto the remnant, changing its mass abruptly, $\Mext\to\Mext+\Delta M$. If
\begin{equation}
\Mext+\Delta M>\Mu,
\label{JumpCondition}
\end{equation}
the system is placed directly on the runaway side of the phase diagram, avoiding the need to grow continuously through the evaporation-dominated interval. Under the chosen evolution law and adequate fuel supply, such deposition can place the system on a growing branch. The condition is dynamical and supply-dependent, rather than a universal remnant mass barrier. A dark-matter core already unstable to collapse can produce a macroscopic black-hole seed without the remnant, with the final mass depending on binding-energy and radiative losses. Its collapse alone therefore does not establish remnant-induced catalysis, although the remnant could affect the collapse history. A distinct nucleation mechanism would require showing that the remnant destabilizes a core that would otherwise remain subcritical. Whether this occurs is an open question.

\subsection{Growth times and critical slowing}
\label{sec:transmutation}

Well above a valid upper threshold, and once hydrodynamics is established, the growth approaches conventional accretion. If Hawking loss is negligible and $K_B$ remains constant, the accretion equation $dM/dt=K_BM^2$ has the solution
\begin{equation}
M(t)
=
\frac{M_i}{1-K_BM_it},
\label{BondiSolution}
\end{equation}
where $M_i\equiv M(t=0)$ is the initial (seed) mass of the black hole at the onset of the hydrodynamic accretion stage, which must lie above the upper threshold, $M_i>\Mu$. The solution has the formal runaway time
\begin{equation}
t_B
\simeq
\frac{1}{K_BM_i}
=
\frac{c_s^3}
{4\pi\lambda\rho_b G^2M_i}
\sim
4\times10^{7}\left(\frac{10^{11}\,\mathrm{kg}}{M_i}\right)\mathrm{yr},
\label{BondiTime}
\end{equation}
where the numerical value assumes $\rho_b=5\times10^{17}\,\mathrm{kg\,m^{-3}}$, $c_s=0.5c$ and $\lambda=1$. The range $8.2\times10^6$--$1.0\times10^8$ years for $c_s=0.3$--$0.7c$ is an extrapolated fixed-background Bondi estimate at $M_i=10^{11}\,\mathrm{kg}$, not a reliable lifetime for a seed in the kinetic crossover. The divergence in Eq.~\eqref{BondiSolution} is not physical: the expanding region influenced by the black hole and the stellar response invalidate a fixed-background treatment before the formal pole. Subsequent consumption or global collapse requires a dynamical stellar calculation~\cite{Richards2021,Baumgarte2021,Schnauck2021}.

Immediately above an unstable balance, Hawking loss cannot be discarded. For the illustrative constant-$K_B$, constant-$\alpha_H$ model, define $x_i=M_i/\Mu>1$ and retain both terms:
\begin{equation}
 \dot M=K_B\frac{M^4-\Mu^4}{M^2},
 \qquad F'(\Mu)=4K_B\Mu>0.
 \label{thresholdDrift}
\end{equation}
Integration to the formal large-mass limit gives
\begin{align}
 t_{\rm run}&=\frac{1}{K_B\Mu}\left[
 \frac14\ln\!\frac{x_i+1}{x_i-1}
 +\frac12\arctan\!\frac1{x_i}\right].
 \label{criticalTime}
\end{align}
For $x_i\gg1$, this reduces to $1/(K_BM_i)$; as $x_i\to1^+$ it diverges logarithmically. The divergence expresses critical slowing near the separatrix. More generally, for a smooth unstable root with slope $s_u=F'(\Mu)>0$, the time to leave a small neighborhood contains $s_u^{-1}\ln[\Delta/(M_i-\Mu)]$, where $\Delta$ is a chosen mass offset. This local statement remains useful when the constant-coefficient Bondi approximation is invalid. Finite stellar supply and global collapse must ultimately replace the formal infinite-mass limit~\cite{Baumgarte2021,Richards2021,Schnauck2021}.

A possible sequence is microscopic capture or relaxation, sufficient sustained or impulsive feeding, growth through a kinetic regime, hydrodynamic accretion, and eventual stellar restructuring. It is not a compulsory sequence for every captured remnant. The conventional stage is insensitive to a Planck-scale $\ell_0$, whereas the time to enter it depends on the microscopic rates and the feeding history.

For the isolated area-scale capture scenario, the initial mass gap and slow capture dominate; for a deposited macroscopic seed, proximity to a kinetic growth threshold and the subsequent accretion law determine the lifetime.

% =================================================
\section{Physical discussion and observational requirements}

\label{sec:discussion}
The central physical distinction is between geometrical regularity, thermal response within a static family, and the evolution of actual quantum states. The metric fixes $\Mext$, the surface-gravity temperature and the far-field correction. It does not fix neutron absorption, the density of remnant states, or decay matrix elements. Positive heat capacity therefore cannot be used as a substitute for an absorption--emission calculation. The canonical lower root is particularly instructive because its excess mass is below the energy resolution of a single capture and its thermal spectrum violates the available energy bound.

There are three broad regimes: a near-ground-state quantum remnant, a small black hole requiring kinetic absorption, and a fluid-accreting black hole with nuclear-scale or larger capture radii. The boundaries between the latter two depend on both wavelength and transport. The first boundary is not a sharp mass demarcation supplied by the static geometry. Table~\ref{tab:scales} distinguishes derived quantities from fiducial estimates and uncertain crossover scales.

\begin{table*}[t]
\centering
\caption{Characteristic scales for $\ell_0=\lPl$, $\rho_b=5\times10^{17}\,\mathrm{kg\,m^{-3}}$ and $v_F=0.4c$. Capture and growth estimates assume the constant-$\mathcal C$ area-scale ansatz and neglect kinetic-energy corrections. The canonical lower root is not a physical equilibrium prediction; the hydrodynamic crossover and upper threshold are model-dependent. Numerical digits are retained for reproducibility, not as a measure of physical accuracy.}
\label{tab:scales}
\begin{tabular}{lll}
\toprule
Quantity & Expression or assumption & Value \\
\midrule
Extremal radius & $\rext=\sqrt2\ell_0$ & $2.29\times10^{-35}\,\mathrm m$ \\
Extremal mass & $\Mext=(3\sqrt3/4)c^2\ell_0/G$ & $2.83\times10^{-8}\,\mathrm{kg}$ \\
Geometrical maximum temperature & Eq.~\eqref{Tmax} & $2.30\times10^{17}\,\mathrm{GeV}/\kB$ \\
Necessary canonical energy scale & Eq.~\eqref{dMstar} & $1.41\times10^{-10}\,\mathrm{kg}$ \\
Fiducial capture rate & $\rho_bv_F\,2\pi\mathcal C\ell_0^2$ & $9.84\times10^{-44}\mathcal C\,\mathrm{kg\,s^{-1}}$ \\
Mean capture interval & Eq.~\eqref{tcap} & $5.39\times10^8\mathcal C^{-1}\,\mathrm{yr}$ \\
Initial nonradiating growth scale & $\Mext/\dot M_q(\Mext)$ & $9.10\times10^{27}\mathcal C^{-1}\,\mathrm{yr}$ \\
Doubling time under the same ansatz & Eq.~\eqref{doublingtime} & $2.06\times10^{27}\mathcal C^{-1}\,\mathrm{yr}$ \\
Formal canonical excess mass & Eq.~\eqref{formalMs} & $1.15\times10^{-45}\sqrt{\mathcal C/g_\ast}\,\mathrm{kg}$ \\
Wavelength comparison & $M_\lambda=c_s^2\bar\lambda_F/G$ & $5.9\times10^{10}$--$3.2\times10^{11}\,\mathrm{kg}$ \\
Illustrative Bondi--Hawking balance & Eq.~\eqref{Mu}, stated parameter range & $3.4\times10^{10}$--$1.3\times10^{11}\,\mathrm{kg}$ \\
Representative kinetic-to-fluid crossover & mean-free-path and wave criteria & $\sim10^{11}$--$10^{12}\,\mathrm{kg}$ \\
\bottomrule
\end{tabular}
\end{table*}

Capture also carries quantum numbers. An angular-momentum scale $J\sim\hbar$ gives $cJ/(G\Mext^2)\sim16/27$ for $\ell_0=\lPl$; using $J=\hbar/2$ as a spin-projection benchmark gives $8/27\simeq0.30$. This is a dimensional comparison, not a Kerr extremality criterion for this regular geometry or an identification of a spin projection with the total quantum angular momentum. Hence a spin-carrying event is not automatically a small perturbation of a nonrotating remnant state, despite its tiny mass increment. The endpoint spectrum as a function of angular momentum and charge, and the allowed compensating emission, must be specified before interpreting a one-neutron event as motion along the neutral static mass curve. Charged near-extremal studies illustrate the role of these selection rules~\cite{Brown2026}; they do not supply the missing state spectrum here.

An excited member of the static family has an inner horizon with nonzero surface gravity. Its persistence under infalling matter, outgoing radiation and interior backreaction is not guaranteed by finite stationary curvature invariants~\cite{Carballo2018,Carballo2021,Carballo2024}. In classical effective descriptions, nondegenerate inner horizons are susceptible to mass inflation. Special constructions with vanishing inner-horizon surface gravity can avoid the corresponding exponential growth~\cite{Carballo2022}, but the $M>\Mext$ members used here do not have that property. The exactly extremal endpoint requires a separate analysis: scalar perturbations of broad classes of degenerate horizons exhibit Aretakis-type nondecay and derivative growth~\cite{Aretakis2015}. These classical results identify stability questions; they do not specify the nonlinear, Planck-scale response of the present effective source. At a Planck-scale core there is additionally no controlled separation between classical curvature and quantum gravity. A phenomenological quasistatic model is useful for isolating assumptions, but statements about a permanently regular remnant must be restricted to that model until its dynamical completion is known.

The neutron-star environment matters for both fluxes. Degenerate occupation factors can modify fermion emission and nuclear collision rates; pairing and superfluidity affect available quasiparticles and transport. Absorption of emitted radiation by the surrounding star does not automatically return that energy to the small hole. Local energy loss, deposited heat and eventual recapture are different contributions and must be tracked separately. A realistic calculation should use stellar temperature, chemical potentials and an equation of state, with a consistent local lapse and a kinetic-to-fluid matching procedure. The regular effective density $\rho_{\rm eff}$ is not the stellar baryon density $\rho_b$.

Old neutron stars can constrain a remnant population only after its production, stellar capture, retention and threshold-crossing probabilities are specified. The existence of a central remnant is the starting assumption here, not a demonstrated outcome of primordial-remnant capture. Extremely weak interaction or gravitational energy loss may make that capture inefficient. Survival constraints also depend on the dark-matter distribution, seed abundance, feeding history and the time needed to reach a destructive regime~\cite{Capela2013,Giddings2008,Adarsha2026}. In the area-scale model, a single remnant captures only a small number of neutrons over a stellar age, so no detectable heating or abundance bound follows from its mere presence. Multiple captured remnants, mergers or an additional interaction would constitute separate growth channels requiring their own rates.

The most direct next calculation is massive-Dirac absorption at the outer horizon, with an explicit extremal limit and the neutron distribution~\cite{Unruh1976,Doran2005}. That should be paired with an energy- and angular-momentum-conserving transition model, and then a stochastic evolution equation for the excitation distribution. For larger masses, the established quantum-accretion calculation and relativistic nuclear-EOS solutions provide benchmarks~\cite{Giffin2022,Schnauck2021}. These steps can determine whether an attracting distribution exists and locate any actual upper growth threshold. They can also assess whether a subcritical dark-matter core is catalyzed by a remnant; collapse of a core already unstable without it is not evidence for such catalysis.

% =================================================
\section{Conclusions}
\label{sec:conclusions}

We have examined a regular microscopic remnant inside neutron-star matter while separating properties of a stationary effective geometry from assumptions about its interaction with the medium. The neutral T-duality-inspired family has a cold endpoint at $\Mext=(3\sqrt3/4)c^2\ell_0/G$ and a near-extremal surface-gravity temperature proportional to $(M-\Mext)^{1/2}$. In a constant-emissivity Stefan--Boltzmann approximation, the power scales as $(M-\Mext)^2$, reaches its maximum after the temperature maximum, and can support a two-root balance structure when the input and loss rates satisfy explicit sign and simplicity conditions.

For $\ell_0=\lPl$, the formal lower canonical root lies far below a single neutron mass in excess energy and outside the canonical domain. It is therefore not a physical continuous equilibrium. A remnant-like capture--relaxation distribution is possible only if allowed microscopic transitions remove the excitation sufficiently effectively; energy conservation and zero temperature alone do not establish this. Angular-momentum transfer, quantum state structure and backreaction are necessary ingredients of that conclusion.

Under the fiducial area-scale absorption assumption, the mean neutron capture interval is $5.4\times10^8\mathcal C^{-1}$ years and the initial nonradiating growth time is $9.1\times10^{27}\mathcal C^{-1}$ years. Thus non-evaporation does not by itself imply rapid stellar destruction. These estimates depend on the absorption ansatz and are not universal predictions of T-duality. Nuclear-scale fluid accretion is separated from the Planck endpoint by roughly nineteen to twenty orders of magnitude in mass, with the crossover determined by transport as well as wavelength. Direct zero-point-length corrections at such radii are negligible.

Sufficient sustained or impulsive feeding can place an object on a growing branch if a valid threshold exists. Subsequent evolution is conventional only after the relevant kinetic and fluid conditions are met, and the growth time is strongly sensitive to proximity to an unstable balance. Old neutron stars therefore probe a combination of capture, absorption, microscopic relaxation and feeding physics, rather than a remnant abundance or a zero-temperature endpoint in isolation. A quantitative prediction requires those ingredients to be calculated within a consistent dynamical theory.

\begin{acknowledgments}
	FSNL acknowledges support from the Funda\c{c}\~ao para a Ci\^encia e a Tecnologia (FCT) research grant UID/04434/2025.
\end{acknowledgments}


\begin{thebibliography}{99}
\interlinepenalty=10000

\bibitem{Capela2013}
	F.~Capela, M.~Pshirkov, and P.~Tinyakov,
	``Constraints on primordial black holes as dark matter candidates
	from capture by neutron stars,''
\href{https://doi.org/10.1103/PhysRevD.87.123524}{Phys.\ Rev.\ D \textbf{87}, 123524 (2013)};
	arXiv:1301.4984 [astro-ph.CO].

\bibitem{Goldman1989}
	I.~Goldman and S.~Nussinov,
	``Weakly interacting massive particles and neutron stars,''
\href{https://doi.org/10.1103/PhysRevD.40.3221}{Phys.\ Rev.\ D \textbf{40}, 3221--3230 (1989)}.

\bibitem{Kouvaris2011}
	C.~Kouvaris and P.~Tinyakov,
	``Constraining asymmetric dark matter through observations
	of compact stars,''
\href{https://doi.org/10.1103/PhysRevD.83.083512}{Phys.\ Rev.\ D \textbf{83}, 083512 (2011)};
	arXiv:1012.2039 [astro-ph.HE].

\bibitem{Richards2021}
	C.~B.~Richards, T.~W.~Baumgarte, and S.~L.~Shapiro,
	``Accretion onto a small black hole at the center of a neutron star,''
\href{https://doi.org/10.1103/PhysRevD.103.104009}{Phys.\ Rev.\ D \textbf{103}, 104009 (2021)};
	arXiv:2102.09574 [astro-ph.HE].

\bibitem{Baumgarte2021}
	T.~W.~Baumgarte and S.~L.~Shapiro,
	``Neutron stars harboring a primordial black hole:
	Maximum survival time,''
\href{https://doi.org/10.1103/PhysRevD.103.L081303}{Phys.\ Rev.\ D \textbf{103}, L081303 (2021)};
	arXiv:2101.12220 [astro-ph.HE].

\bibitem{RichardsBondi2021}
	C.~B.~Richards, T.~W.~Baumgarte, and S.~L.~Shapiro,
	``Relativistic Bondi accretion for stiff equations of state,''
\href{https://doi.org/10.1093/mnras/stab161}{Mon.\ Not.\ R.\ Astron.\ Soc.\ \textbf{502}, 3003--3011 (2021)};
	arXiv:2101.08797 [astro-ph.HE].

\bibitem{RichardsErratum2021}
	C.~B.~Richards, T.~W.~Baumgarte, and S.~L.~Shapiro,
	``Erratum: Relativistic Bondi accretion for stiff equations of state,''
\href{https://doi.org/10.1093/mnras/stab2069}{Mon.\ Not.\ R.\ Astron.\ Soc.\ \textbf{506}, 3935 (2021)}.

\bibitem{Schnauck2021}
	S.~C.~Schnauck, T.~W.~Baumgarte, and S.~L.~Shapiro,
	``Accretion onto black holes inside neutron stars with
	piecewise-polytropic equations of state:
	Analytic and numerical treatments,''
\href{https://doi.org/10.1103/PhysRevD.104.123021}{Phys.\ Rev.\ D \textbf{104}, 123021 (2021)}.

\bibitem{Giffin2022}
	P.~Giffin, J.~Lloyd, S.~D.~McDermott, and S.~Profumo,
	``Neutron star quantum death by small black holes,''
\href{https://doi.org/10.1103/PhysRevD.105.123030}{Phys.\ Rev.\ D \textbf{105}, 123030 (2022)};
	arXiv:2105.06504 [hep-ph].

\bibitem{Adarsha2026}
	H.~A.~Adarsha, C.~Chakraborty, and S.~Bhattacharyya,
	``Transmutation timescales for the dark matter induced collapse
	of compact stars into black holes,''
\href{https://doi.org/10.1103/zb1m-762n}{Phys.\ Rev.\ D \textbf{114}, 063007 (2026)};
	arXiv:2608.10594 [astro-ph.HE].

\bibitem{Hawking1975}
	S.~W.~Hawking,
	``Particle creation by black holes,''
\href{https://doi.org/10.1007/BF02345020}{Commun.\ Math.\ Phys.\ \textbf{43}, 199--220 (1975)}.

\bibitem{Padmanabhan1997}
	T.~Padmanabhan,
	``Duality and zero-point length of spacetime,''
\href{https://doi.org/10.1103/PhysRevLett.78.1854}{Phys.\ Rev.\ Lett.\ \textbf{78}, 1854--1857 (1997)};
	arXiv:hep-th/9608182.

\bibitem{Fontanini2006}
	M.~Fontanini, E.~Spallucci, and T.~Padmanabhan,
	``Zero-point length from string fluctuations,''
\href{https://doi.org/10.1016/j.physletb.2005.12.039}{Phys.\ Lett.\ B \textbf{633}, 627--630 (2006)};
	arXiv:hep-th/0509090.

\bibitem{Nicolini2019}
	P.~Nicolini, E.~Spallucci, and M.~F.~Wondrak,
	``Quantum corrected black holes from string T-duality,''
\href{https://doi.org/10.1016/j.physletb.2019.134888}{Phys.\ Lett.\ B \textbf{797}, 134888 (2019)};
	arXiv:1902.11242 [gr-qc].

\bibitem{Jusufi:2024dtr}
	K.~Jusufi and P.~Nicolini,
	``Geodesic completeness from string T-duality,''
\href{https://doi.org/10.1140/epjc/s10052-025-15018-8}{Eur.\ Phys.\ J.\ C \textbf{85}, 1291 (2025)};
	arXiv:2410.19613 [hep-th].

\bibitem{Jusufi:2026jan}
	K.~Jusufi and F.~S.~N.~Lobo,
	``Quantum-gravity-inspired Alcubierre warp-drive geometries,''
	\href{https://arxiv.org/abs/2609.05554}{arXiv:2609.05554} [gr-qc] (2026).

\bibitem{Chen2015}
	P.~Chen, Y.~C.~Ong, and D.-h.~Yeom,
	``Black hole remnants and the information loss paradox,''
\href{https://doi.org/10.1016/j.physrep.2015.10.007}{Phys.\ Rept.\ \textbf{603}, 1--45 (2015)};
	arXiv:1412.8366 [gr-qc].

\bibitem{Dymnikova2015}
	I.~Dymnikova and M.~Khlopov,
	``Regular black hole remnants and graviatoms with de Sitter
	interior as heavy dark matter candidates probing inhomogeneity
	of early universe,''
\href{https://doi.org/10.1142/S0218271815450029}{Int.\ J.\ Mod.\ Phys.\ D \textbf{24}, 1545002 (2015)};
	arXiv:1510.01351 [gr-qc].

\bibitem{Jusufi:2025qgd}
	K.~Jusufi and D.~Singleton,
	``Regular black holes with gravitational self-energy as dark matter,''
\href{https://doi.org/10.1140/epjc/s10052-026-15787-w}{Eur.\ Phys.\ J.\ C \textbf{86}, 530 (2026)};
	arXiv:2509.13335 [gr-qc].

\bibitem{Giddings2008}
	S.~B.~Giddings and M.~L.~Mangano,
	``Astrophysical implications of hypothetical stable
	TeV-scale black holes,''
\href{https://doi.org/10.1103/PhysRevD.78.035009}{Phys.\ Rev.\ D \textbf{78}, 035009 (2008)};
	arXiv:0806.3381 [hep-ph].

\bibitem{Lobo:2025nng}
F.~S.~N.~Lobo and M.~E.~Rodrigues,
``Quantum-inspired wormholes from string T-duality,''
\href{https://doi.org/10.1103/wg37-gg1l}{Phys.\ Rev.\ D \textbf{112}, 064082 (2025)};
arXiv:2506.17950 [gr-qc].

\bibitem{Lobo:2026dft}
F.~S.~N.~Lobo and M.~E.~Rodrigues,
``Linearized stability of T-duality quantum-inspired
thin-shell wormholes,''
\href{https://doi.org/10.1103/hr5g-nnxr}{Phys.\ Rev.\ D \textbf{114}, 064086 (2026)};
arXiv:2606.11413 [gr-qc].

\bibitem{Chen2026}
	C.-M.~Chen, K.~Jusufi, and D.~Singleton,
	``Schwinger stability of T-duality-inspired extremal black holes,''
	\href{https://arxiv.org/abs/2609.18528}{arXiv:2609.18528} [hep-th] (2026).

\bibitem{Carballo2018}
	R.~Carballo-Rubio, F.~Di~Filippo, S.~Liberati,
	C.~Pacilio, and M.~Visser,
	``On the viability of regular black holes,''
\href{https://doi.org/10.1007/JHEP07(2018)023}{J.\ High Energy Phys.\ \textbf{07} (2018), 023};
	arXiv:1805.02675 [gr-qc].

\bibitem{Page1976}
	D.~N.~Page,
	``Particle emission rates from a black hole:
	Massless particles from an uncharged, nonrotating hole,''
\href{https://doi.org/10.1103/PhysRevD.13.198}{Phys.\ Rev.\ D \textbf{13}, 198--206 (1976)}.

\bibitem{Bai2023}
	Y.~Bai and M.~Korwar,
	``Near-extremal charged black holes: Greybody factors and evolution,''
\href{https://doi.org/10.1007/JHEP03(2023)151}{J.\ High Energy Phys.\ \textbf{03} (2023), 151};
	arXiv:2301.07739 [hep-th].

\bibitem{Shafiee2026}
M.~Shafiee and A.~Sheykhi,
``Greybody factors and Hawking spectra of quantum-corrected black holes,''
\href{https://arxiv.org/abs/2609.32702}{arXiv:2609.32702} [hep-th] (2026).

\bibitem{Preskill1991}
	J.~Preskill, P.~Schwarz, A.~Shapere, S.~Trivedi, and F.~Wilczek,
	``Limitations on the statistical description of black holes,''
\href{https://doi.org/10.1142/S0217732391002773}{Mod.\ Phys.\ Lett.\ A \textbf{6}, 2353--2362 (1991)}.

\bibitem{Casadio1998}
	R.~Casadio and B.~Harms,
	``Microfield dynamics of black holes,''
\href{https://doi.org/10.1103/PhysRevD.58.044014}{Phys.\ Rev.\ D \textbf{58}, 044014 (1998)}.

\bibitem{Parikh2000}
	M.~K.~Parikh and F.~Wilczek,
	``Hawking radiation as tunneling,''
\href{https://doi.org/10.1103/PhysRevLett.85.5042}{Phys.\ Rev.\ Lett.\ \textbf{85}, 5042--5045 (2000)};
	arXiv:hep-th/9907001.

\bibitem{Brown2026}
	A.~R.~Brown, L.~V.~Iliesiu, G.~Penington, and M.~Usatyuk,
	``The evaporation of charged black holes,''
\href{https://doi.org/10.1007/JHEP01(2026)109}{J.\ High Energy Phys.\ \textbf{01} (2026), 109};
	arXiv:2411.03447 [hep-th].

\bibitem{Das1997}
	S.~R.~Das, G.~W.~Gibbons, and S.~D.~Mathur,
	``Universality of low energy absorption cross sections
	for black holes,''
\href{https://doi.org/10.1103/PhysRevLett.78.417}{Phys.\ Rev.\ Lett.\ \textbf{78}, 417--419 (1997)};
	arXiv:hep-th/9609052.

\bibitem{Unruh1976}
	W.~G.~Unruh,
	``Absorption cross section of small black holes,''
\href{https://doi.org/10.1103/PhysRevD.14.3251}{Phys.\ Rev.\ D \textbf{14}, 3251--3259 (1976)}.

\bibitem{Doran2005}
	C.~Doran, A.~Lasenby, S.~Dolan, and I.~Hinder,
	``Fermion absorption cross section of a Schwarzschild black hole,''
\href{https://doi.org/10.1103/PhysRevD.71.124020}{Phys.\ Rev.\ D \textbf{71}, 124020 (2005)};
	arXiv:gr-qc/0503019.

\bibitem{Emparan2025}
	R.~Emparan,
	``Quantum cross-section of near-extremal black holes,''
\href{https://doi.org/10.1007/JHEP04(2025)122}{J.\ High Energy Phys.\ \textbf{04} (2025), 122};
	arXiv:2501.17470 [hep-th].

\bibitem{Bondi1952}
	H.~Bondi,
	``On spherically symmetrical accretion,''
\href{https://doi.org/10.1093/mnras/112.2.195}{Mon.\ Not.\ R.\ Astron.\ Soc.\ \textbf{112}, 195--204 (1952)}.

\bibitem{Michel1972}
	F.~C.~Michel,
	``Accretion of matter by condensed objects,''
\href{https://doi.org/10.1007/BF00649949}{Astrophys.\ Space Sci.\ \textbf{15}, 153--160 (1972)}.

\bibitem{MacGibbon1991}
	J.~H.~MacGibbon,
	``Quark- and gluon-jet emission from primordial black holes.
	II. The emission over the black-hole lifetime,''
\href{https://doi.org/10.1103/PhysRevD.44.376}{Phys.\ Rev.\ D \textbf{44}, 376--392 (1991)}.

\bibitem{Carballo2021}
R.~Carballo-Rubio, F.~Di~Filippo, S.~Liberati, C.~Pacilio, and M.~Visser,
``Inner horizon instability and the unstable cores of regular black holes,''
\href{https://doi.org/10.1007/JHEP05(2021)132}{J.\ High Energy Phys.\ \textbf{05} (2021), 132};
arXiv:2101.05006 [gr-qc].

\bibitem{Carballo2024}
R.~Carballo-Rubio, F.~Di~Filippo, S.~Liberati, and M.~Visser,
``Mass inflation without Cauchy horizons,''
\href{https://doi.org/10.1103/PhysRevLett.133.181402}{Phys.\ Rev.\ Lett.\ \textbf{133}, 181402 (2024)};
arXiv:2402.14913 [gr-qc].

\bibitem{Carballo2022}
R.~Carballo-Rubio, F.~Di~Filippo, S.~Liberati, C.~Pacilio, and M.~Visser,
``Regular black holes without mass inflation instability,''
\href{https://doi.org/10.1007/JHEP09(2022)118}{J.\ High Energy Phys.\ \textbf{09} (2022), 118};
arXiv:2205.13556 [gr-qc].

\bibitem{Aretakis2015}
S.~Aretakis,
``Horizon instability of extremal black holes,''
\href{https://doi.org/10.4310/ATMP.2015.v19.n3.a1}{Adv.\ Theor.\ Math.\ Phys.\ \textbf{19}, 507--530 (2015)};
arXiv:1206.6598 [gr-qc].

\end{thebibliography}
\end{document}